\documentclass[onecolumn,aps,pre,groupedaddress,showpacs,floatfix,preview]{revtex4-1}
\usepackage{graphicx}
\usepackage{epsfig}
\usepackage{amsfonts}
\usepackage{color}
\usepackage{ulem}
\usepackage{amsmath}
\usepackage{mathrsfs}
\usepackage{subfigure}
\usepackage[autostyle]{csquotes}

\begin{document}

\title{Novel Universality Classes in Ferroelectric Liquid Crystals}

\author{Amit K Chattopadhyay}
\affiliation{
Aston University, System Analytics Research Institute, Birmingham, B4 7ET, UK}
\email{a.k.chattopadhyay@aston.ac.uk}
\author{Prabir K. Mukherjee}
\affiliation{Department of Physics, Government College of Engineering
and Textile Technology, 12 William Carey Road, Serampore, Hooghly-712201,
India}
%\email{pkmuk1966@gmail.com}

\begin{abstract}
Starting from a Langevin formulation of a thermally perturbed nonlinear 
elastic model of the ferroelectric smectic-C$^*$ (SmC${*}$) liquid crystals in the presence of 
an electric field, this article characterizes the hitherto unexplored 
dynamical phase transition from a thermo-electrically forced ferroelectric 
SmC${}^{*}$ phase to a chiral nematic liquid crystalline phase and vice versa. 
{{The theoretical analysis is based on a combination of dynamic renormalization 
(DRG) and numerical simulation of the emergent model.}} While the DRG architecture predicts a 
generic transition to the Kardar-Parisi-Zhang (KPZ) universality class at 
dynamic equilibrium, in agreement with recent experiments, the numerical simulations of the model show simultaneous 
existence of two phases, one a \enquote{subdiffusive} (SD) phase characterized 
by a dynamical exponent value of 1, and the other a KPZ phase, characterized 
by a dynamical exponent value of 1.5. The SD phase flows over to the KPZ phase 
with increased external forcing, offering a new universality paradigm, hitherto unexplored in the context of ferroelectric liquid crystals. 
\end{abstract}

\maketitle

\newpage
%\doublespacing
\newpage

\section*{Introduction}

Kardar-Parisi-Zhang (KPZ) equation \cite{kpz, barabasi, halpinhealy, krug} has long been proposed as a paradigmatic model characterizing the statistical properties of a growing interface. While the model itself is a pure theorist's delight, in that it offers both weak and strong coupling regimes with transitions in between, a field theoretic paradise of sorts, from an experimental perspective, the model has often been looked down upon apart from rare conjectures and interpolations \cite{myllys1, myllys2}. 

On the other hand, researches on smectic-A to SmC$^*$ (SmA-SmC$^*$) phase transitions
have been going on over the past few decades. Now there is also an increasing interest in the chiral nematic (N$^*$) to SmC$^*$ (N$^*$-SmC$^*$)  phase transitions, both because of their inherent challenge \cite{hiller} as also due to the application potentials \cite{kumar}. These studies have generally relied on deterministic free energy based descriptions and follow-up estimations without much importance ascribed to the inherent thermal or boundary fluctuations surrounding the samples concerned. When such thermally forced stochastic perturbations are considered, the liquid crystalline phases may indeed transgress their conventional regimes and cross over to hitherto unexplored (\enquote{hidden}) regimes, as recently shown in \cite{Chattopadhyay2015}. I

{{The present work studies the impact of stochastic
fluctuations in modifying the spatiotemporal properties of cholesteric liquid crystals in the presence of a random 
magnetic field. Our results show that stochastic  fluctuations drive the system to a Kosterlitz-Thouless 
transition point, through a second-order phase transition, thereby converging to a Kardar-Parisi-Zhang universality
class. For weaker ramping forces, the system converges to a non-KPZ fixed point.}}

Such mapping to established universality classes is unlikely to be restricted to only one or two specific cases, which is why there has been an upsurge in interest in stochastic modeling from the perspective of analyzing some dynamical propagation fronts and related crossovers and/or phase transitions in liquid crystalline systems
\cite{Chattopadhyay2015,sano1,sano,wang1,wang2,geng}. Such phase transitions range from geometric phases \cite{sergei} to isotropic-nematic-columnar phases \cite{caroline, felix}. Nematic liquid crystal
turbulence has been shown to exhibit a clear Kardar-Parisi-Zhang (KPZ)-class behavior 
\cite{sano1,sano}. Takeuchi and Sano \cite{sano1,sano} experimentally studied the 
scale-invariant fluctuations of growing
interfaces in nematic liquid-crystal turbulence.  
They observed that the interfaces exhibit self-affine roughening characterized by both spatial and temporal
scaling laws of the KPZ theory in 1+1 dimensions. Golubovic and Wang \cite{wang1,wang2} found a theoretical relationship between the fluctuations of smectic-A 
to fluctuations KPZ dynamical model. They observed that the KPZ model in 2+1 
dimensions maps into a elastic critical point of 3D Smectic-A with broken 
inversion symmetry (ferroelectric smectic-A). 

Here we report on what we believe to be the first theoretical study of 
the dynamical evolution patterns of ferroelectric smectic-C$^*$ (SmC$^*$) liquid crystals in an electric 
field, using firstly the language of dynamic renormalization group (DRG), which we later complement with numerical
simulation of the propounded stochastic model. {{Both analytical (DRG based) and numerical (simulation) results predict a chiral nematic (N$^*$) to  SmC$^*$ phase transition.}}
In the experimentally viable low frequency limit, our 
dynamical model predicts the onset of a Kosterlitz-Thouless transition. At large enough spatiotemporal scales, the inherent nonlinearity drives the system to a 
Kardar-Parisi-Zhang fixed point, theoretically represented by appropriate self-affine scaling.
  
\section*{The Theoretical Model}
\label{theoretical_model}

The inner fabric of our proposed model relies on the nonlinear elastic model 
of SmC$^*$ liquid crystal in the
presence of an electric field. {{The SmC$^*$ phase has a helical structure
where the director field tilts from the layer normal by $\theta$ and rotates 
along normal direction. An electric polarization appears in the plane 
parallel with the layer and makes a right angle along the director field and 
in the result it rotates along the layer normal. By applying the electric 
field in a direction parallel to the layer, the helical structure is unwound
due to a coupling between the electric field and the polarization. 
We assume all elastic constants equal, define a two
dimensional director in the plane of the layers.
So the the director $\bf n$ is defined as:
$n_x=\sin\theta\cos\phi(z)$, $n_y=\sin\theta\sin\phi(z)$. $n_z=\cos\theta$. $\phi(z)$ is the azimuthal angle
between $\bf n$ and y axis and is thus z dependent. $\theta$ is the tilt angle between the layer 
normal and the
director $\bf n$ and does not vary with z. Here layer normal is assumed to be 
in the z direction.}}
In what follows, we first define a Frank free energy, as shown in Eq. (\ref{free1}) below. This is then followed by a Langevin formulation, starting from this free energy, to model the time dynamics of this system. We shall restrict our
interest to a one dimensional dynamical problem in space. 

\subsection*{The Frank Free Energy}
%\label{frank_free_energy}
Based on the Cladis-Saarloos formalism \cite{cladis},  
the free energy of the SmC$^*$ phase in an applied electric field ${\bf E}=E(0,1,0)$ can be written as

\begin{equation}
g=\int\left[\frac {K\theta^2}{2}\left(\frac {d \phi}{dz}+q_0\right)^2
-\frac {\epsilon_a\theta^2}{8\pi}E^2\sin^2\phi-P\theta E\cos\phi\right]dz,
\label{free1}
\end{equation}
where $K$ is the twist elastic constant, $q_0$ the wave vector corresponding
to $E=0$ ($q_0=2\pi/\lambda_0)$. $\epsilon_a$ is the liquid crystal dielectric
anisotropy. $P$ is the polarization. 
We assume polarization $P$ rotates about the z axis and the electric field is applied in the 
y direction. The 
above free energy (\ref{free1}) was extensively studied by Cladis and 
Saarloos \cite{cladis} to discuss the various the forms of front propagation
of the SmC$^*$ state. The functional integral shown in Eq. (\ref{free1}) above can be approximately solved in certain cases using a saddle point approximation method but this will not work in the non-asymptotic regime, the focal point of our interest \cite{modanese2006}.

\subsection*{The Langevin Model}
%\label{langevin_model}

The free energy defined in Eq. (\ref{free1}) can be used to arrive at the stochastic Langevin description, as presented in Risken \cite{Risken}. The time dependent Ginzburg-Landau (TDGL)-type model can then be specified by the following Langevin dynamics

\begin{equation}
M\frac {\partial \phi(z,t)}{\partial t}=-\frac {\delta g}{\delta \phi} + F_0
+\eta(z,t),
\label{lang1}
\end{equation}
where $\eta(z,t)$ is a stochastic white noise represented by $<\eta(z,t)\: 
\eta(z',t')> = D_0 \:\delta(z-z') \delta(t-t')$ and $<\eta(z,t)>=0$, 
in which the curly brackets \enquote{$< >$} represent ensemble average over 
all noise realizations and $D_0$ is the noise strength, a constant, while 
$\frac{\delta g}{\delta \phi}$ is the functional derivative of $g$ with 
respect to variable $\phi$. {{$F_0$ is a constant “overdamped
force” that pumps a steady energy into the system.
Such a term is related to the existence of a 
kinetic roughening in the model that can only be counter-
balanced by a negative damping term $F_0$.}}

Eq. (\ref{lang1}) can then be rewritten as \\

\begin{equation}
M\frac {\partial \phi}{\partial t}=\frac {\epsilon_a\theta^2E^2}{8\pi}\sin2\phi
-P\theta E \sin\phi+K
\theta^2\frac {\partial^2 \phi}{\partial z^2}
+F_0+\eta(z,t).
\label{lang2}
\end{equation}

The above Eq. (\ref{lang2}) represents a quintessential sine-Gordon model \cite{nozieres} whose low frequency spatiotemporal scaling properties have been studied previously by Chattopadhyay \cite{akc}, an approach that was later implemented in analyzing a class of liquid crystals by Chattopadhyay and Mukherjee \cite{Chattopadhyay2015}.

We will now calculate the renormalization group (RG) flows for the dynamical
nonlinear model presented in Eq. (\ref{lang2}) above. In order to achieve a scale independent formulation of the aforementioned dynamics, in the following, we rescale the starting model variable on a periodic lattice using the transformation $h=\frac{a}{2\pi}\phi$, which 
then leads to the following representation:

\begin{equation}
\eta\frac {\partial h}{\partial t}=\gamma \frac {\partial^2h}{\partial z^2}-\frac {2\pi V_1}
{a}\sin\left(\frac {2\pi h}{a}\right)+\frac {2\pi V_2}{a}\sin\left(\frac {4\pi h}{a}\right)+F+N(z,t).
\label{lang5}
\end{equation}

Here the rescaled parameters are given by $\eta=\frac {M}{K}$, $\gamma=\theta^2$, $\frac {2\pi V_1}{a}=\frac
{P\theta Ea}{2\pi K}$, $\frac {2\pi V_2}{a}=\frac {\epsilon_a\theta^2E^2a}{16\pi^2 
K}$, $F=\frac{F_0a}{2\pi K}$ and the rescaled noise $N(z,t)=\frac {a\eta^{\prime}(z,t)}
{2\pi K}$. The model above is a generalized extension of the classic noisy sine-Gordon  
model that can be studied using the method previously used by
Chattopadhyay \cite{akc} and later contextualized for liquid crystals in \cite{Chattopadhyay2015}, {{in which the second moment of the noise-noise correlation in the $k$-space can be given by
\begin{equation}
<N(k,t) N(k',t') = 2D f\left(\dfrac{\pi k}{\Lambda}\right)\:\delta(k+k')\:\delta(t-t'),
\label{noiseeq}
\end{equation}

where $f(1-x)=\theta(x)$, a Heaviside step function, $\Lambda \propto \frac{\pi}{a}$, and the rescaled noise strength $D=D_0\frac{a^2}{4\pi^2 K^2}$ is defined through the fluctuation-dissipation theorem \cite{Derrida2007} as $D \propto \eta T$, where $T$ is the \enquote{non-equilibrium temperature} \cite{Jarzynski2011}}}. 

\section*{Results}
Eq. (\ref{lang5}) defines a stochastically evolving structure which we solve independently using two well established architectures - first, using the Dynamic Renormalization Group (DRG) approach and second, a thorough numerical solution of the {{same stochastic model using appropriate discretization}} that simultaneously offers a complementary strand to the DRG method while also probing regimes that go beyond the perturbative mechanism accorded under the DRG auspice. {{Note that ours being a stochastically perturbed 1+1 dimensional model, it is open to phase transition possibilities.}}

\subsection*{The Dynamic Renormalization Group Analysis}
\label{DRG}

The model presented in Eqs. (\ref{lang5}-\ref{noiseeq}) is the d=1+1 dimensional version of the
Nozieres-Gallet \cite{nozieres} or Rost-Spohn model \cite{rost} for the special case of
$\lambda=0$. In the following analysis, we will employ the same dynamic renormalization group 
(DRG) prescription as in  \cite{rost}. Our effective dynamical equation for the description is the following:

\begin{equation}
\eta\frac {\partial h}{\partial t}=\gamma \frac {\partial^2h}{\partial z^2}-\frac {2\pi V_1}
{a}\sin\left(\frac {2\pi h}{a}\right)+\frac {2\pi V_2}{a}\sin\left(\frac {4\pi h}{a}\right)+F+\frac{\lambda}{2} {\bigg( \dfrac{\partial h}{\partial z} \bigg)}^2 + N(z,t),
\label{DRGeqn}
\end{equation}

where the KPZ nonlinearity $\dfrac{\lambda}{2}{\bigg( \dfrac{\partial h}{\partial z} \bigg)}^2$ has been added in advance, in recognition of
the same term appearing after the renormalization, as studied in \cite{rost,Chattopadhyay2015}. All other terms bear the same meaning as in Eq. (\ref{lang5}) before.

The fast frequency components are integrated over the 
momentum shell $\Lambda e^{-\Delta l}<|k|<\Lambda$ 
i.e. $\Lambda (1-\Delta l)$ for $<|k|<\Lambda$, in the limit of 
infinitesimally small $\Delta l$. We assume both $V_1$ and $V_2$ are perturbative 
constants. 
%Now the effective free energy functional for zero driving force ($F=0$ and $N=0$) can be written as 

%\begin{equation}
%g=\int dz \Big[ \frac {\gamma}{2}{\bigg(\frac{\partial h}{\partial z}\bigg)}^2-V\cos(\frac {2\pi}{a}h) \Big],
%\label{free3}
%\end{equation}
Within this momentum shell, the variables are renormalized as follows: $k\rightarrow k^{\prime}=
(1+\Delta l)k$, $z\rightarrow z^{\prime}=(1-\Delta l)z$, $h\rightarrow h^{\prime}=h$,
$t\rightarrow t^{\prime}=(1-2 \Delta l)t$, $\eta\rightarrow \eta^{\prime}=\eta$,
$\gamma\rightarrow \gamma^{\prime}=\gamma$, 
$V_1\rightarrow V_1^{\prime}=(1+2 \Delta l)V_1$, 
$V_2\rightarrow V_2^{\prime}=(1+2 \Delta l)V_2$, $F\rightarrow F^{\prime}=(1+2 \Delta l)F$ and $\lambda \to \lambda'=\lambda$.

First we discuss the perturbative dynamics. In this case Eq. (\ref{DRGeqn}) 
can be rewritten as
\begin{equation}
\eta\frac {\partial h}{\partial t}=\gamma \frac{\partial^2 h}{\partial z^2}+\Psi(h)+N,
\label{lang6}
\end{equation}
where $\Psi(h)=-\frac {2\pi V_1}{a}\sin[\frac {2\pi}{a}(h+\frac{Ft}{\eta})]
+\frac {2\pi V_2}{a}\sin[\frac {4\pi}{a}(h+\frac{Ft}{\eta})] + \frac{\lambda}{2} {\big( \frac{\partial h}{\partial z} \big)}^2$.

Using perturbation expansions for the dynamic variables $X_i=h, N$, we can 
write $N=\bar{N}+\delta N$, where the quantity $\bar{N}$ is defined within the momentum range $|k|<(1-\Delta l)\Lambda$, with $\delta N$ defined inside the annular ring 
$(1-\Delta l)\Lambda<|k|<\Lambda$, to get

\begin{eqnarray} 
\eta\frac {\partial \bar{h}}{\partial t}&=&\gamma \frac{\partial^2 \bar{h}}{\partial z^2}+\bar{\Psi}(\bar{h},\delta h)+\bar{N},\nonumber \\
&&\eta\frac {\partial \delta h}{\partial t}=\gamma \frac{\partial^2 \delta h}{\partial z^2}+\delta \Psi(\bar{h},\delta h)+\delta N,
\label{lang7}
\end{eqnarray}
where $\bar{\Psi}=<\Psi>_{\delta N}$, averaging defined over all noise perturbations.
$\bar{\Psi}$ and $\delta {\Psi}$ can be expressed as 

\begin{eqnarray}
\bar{\Psi}&=&-\frac {2\pi V_1}{a}\sin \bigg[\frac {2\pi}{a}(\bar{h}+\frac{F t}{\eta})\bigg]
\left(1-\frac {2\pi^2}{a^2}<\delta h^2>_{\delta N}\right)
+\frac {2\pi V_2}{a}\sin \bigg[\frac {4\pi}{a}(\bar{h}+\frac{F t}{\eta})\bigg]
\left(1-\frac {8\pi^2}{a^2}<\delta h^2>_{\delta N}\right) \nonumber \\
&+& \frac{\lambda}{2}{\bigg(\frac{\partial {\bar h}}{\partial z}\bigg)}^2 +  \frac{\lambda}{2}<{\bigg(\frac{\partial {\delta h}}{\partial z}\bigg)}^2>_{\delta N}
\nonumber \\
&&\delta {\Psi}=\bigg\{-\frac {4\pi^2 V_1}{a^2}\cos \bigg[\frac {2\pi}{a}(\bar{h}+\frac{F t}{\eta})\bigg]+\frac {8\pi^2 V_2}{a^2}
\cos\bigg[\frac {4\pi}{a}\left(\bar{h}\frac{F t}{\eta}\right)\bigg] \bigg\} \delta h + \lambda \bigg(\dfrac{\partial \bar h}{\partial z}\bigg) .\bigg(\dfrac{\partial \delta h}{\partial z}\bigg).
\label{npsi1}
\end{eqnarray}

Expanding $\delta h=\delta h^{(0)}+\delta h^{(1)}+....$ perturbatively, {{in the Fourier transformed $k$-space,}} we get
{{
%\begin{subequations}
\begin{equation}
\delta h^{(0)}(z,t) =\int^t_{-\infty}dt^{\prime}\int dz^{\prime}\: G_0(z-z^{\prime},
t-t^{\prime})\:\delta N(z^{\prime},t^{\prime})
\label{expand1}
\end{equation}
and
\begin{eqnarray}
\delta h^{(1)}(z,t)&=&\int^t_{-\infty}dt^{\prime}\int dz^{\prime}\: G_0(z-z^{\prime},
t-t^{\prime})\times\bigg[-\frac {4\pi^2 V_1}{a^2}\cos\bigg(\frac {2\pi}{a}(\bar{h}
(z^{\prime},t^{\prime})+\frac{Ft^{\prime}}{\eta}\bigg) \nonumber \\
&+&\frac {8\pi^2 V_2}{a^2}
\cos\bigg(\frac {4\pi}{a}(\bar{h}(z^{\prime},t^{\prime})+\frac{Ft^{\prime}}{\eta}\bigg)\bigg]\: \delta h^{(0)} + \lambda \bigg(\frac{\partial \bar h(z',t')}{\partial z'}\bigg) \bigg(\frac{\partial \delta h^{(0)}(z',t')}{\partial z'}\bigg),
\label{expand2}
\end{eqnarray}
%\end{subequations}

where $G_0(x,t)=\frac {1}{2\pi \gamma t}e^{-\frac {\eta x^2}{2\gamma t}}$ is 
the Green's function. For the initial condition $h(z,t=-\infty)=0$, we get 
\begin{equation}
<{(\delta h)}^2(z,t)>=<{(\delta h^{(0)})}^2>+2<\delta h^{(0)}(z,t)\:
\delta h^{(1)}(z,t)>
\label{expand3}
\end{equation}
and
\begin{equation}
<(\partial_z \delta h)^2(z,t)> = <(\partial_z \delta h^{(0)})^2>+2<\partial_z \delta
h^{(0)}\: \partial_z \delta h^{(1)}>.
\label{expand4}
\end{equation}
%\end{subequations}

Using Eqs. (\ref{expand1}-\ref{expand4}), we can now write}}
\begin{equation}
\delta h_k(t)=\frac {1}{\eta}e^{-\frac {\gamma}{\eta} k^2 t}\int_{-\infty}^tdt^{\prime}
\delta N_k(t^{\prime})e^{\frac {\gamma}{\eta} k^2 t^{\prime}}
\label{lang9}
\end{equation}
The corresponding correlation functions are given by
\begin{equation}
\bigg \langle \delta h^{(0)}(z,t)\:\delta h^{(0)}(z^{\prime},t^{\prime}) \bigg\rangle=\frac {D}
{\gamma\eta}e^{i\Lambda (z-z^{\prime})}\:e^{-\frac {\gamma}{\eta}\Lambda^2
|t-t^{\prime}|} \Delta l\:\:\text{and}
\label{corre3}
\end{equation}
\begin{equation}
\bigg \langle \bigg(\frac{\partial \delta h^{(0)}(z,t)}{\partial z}\bigg)\:\bigg(\frac{\partial \delta h^{(0)}(z^{\prime},t^{\prime})}{\partial z'}\bigg) \bigg \rangle=-\frac {D}
{\gamma \eta}\Lambda^2e^{i\Lambda (z-z^{\prime})}\:e^{-\frac {\gamma}{\eta}\Lambda^2
|t-t^{\prime}|}\Delta l {\bf \:\:},
\label{corre4}
\end{equation}
%\end{subequations}
where $D=\frac {a^2}{4\pi^2K^2}\eta T$.

\subsection*{Dynamic Renormalization Group Flow Equations}
The Dynamic Renormalization Group (DRG) flow equations, correct up to the first order, can be shown to be as follows:
%\begin{subequations}
\begin{equation}
dV_1^{(l)}=-\frac {2\pi^2}{a^2}<\delta h^{(0)2}(z,t)>=-\frac {\eta T}{2\gamma 
\eta K^2}\Delta l,
\label{first1}
\end{equation}
\begin{equation}
dV_2^{(l)}=-\frac {8\pi^2}{a^2}<\delta h^{(0)2}(z,t)>=-\frac {2\eta T}{\gamma 
\eta K^2}\Delta l,
\label{first2}
\end{equation}
\begin{equation}
dF^{(l)}=\dfrac{\lambda}{2}<{\bigg( \frac{\partial \delta h^{(0)}}{\partial z}\bigg)}^2>_{\delta N}=\dfrac{\lambda T}{2\pi \gamma} \Lambda^2 \Delta l.
\label{first3}
\end{equation}
%\end{subequations}

We now obtain the first order dynamic renormalization group (DRG) flows for $V$ and $F$. 
\begin{center}
\begin{figure}[ht]
\includegraphics[height=10.0cm,width=12.0cm]{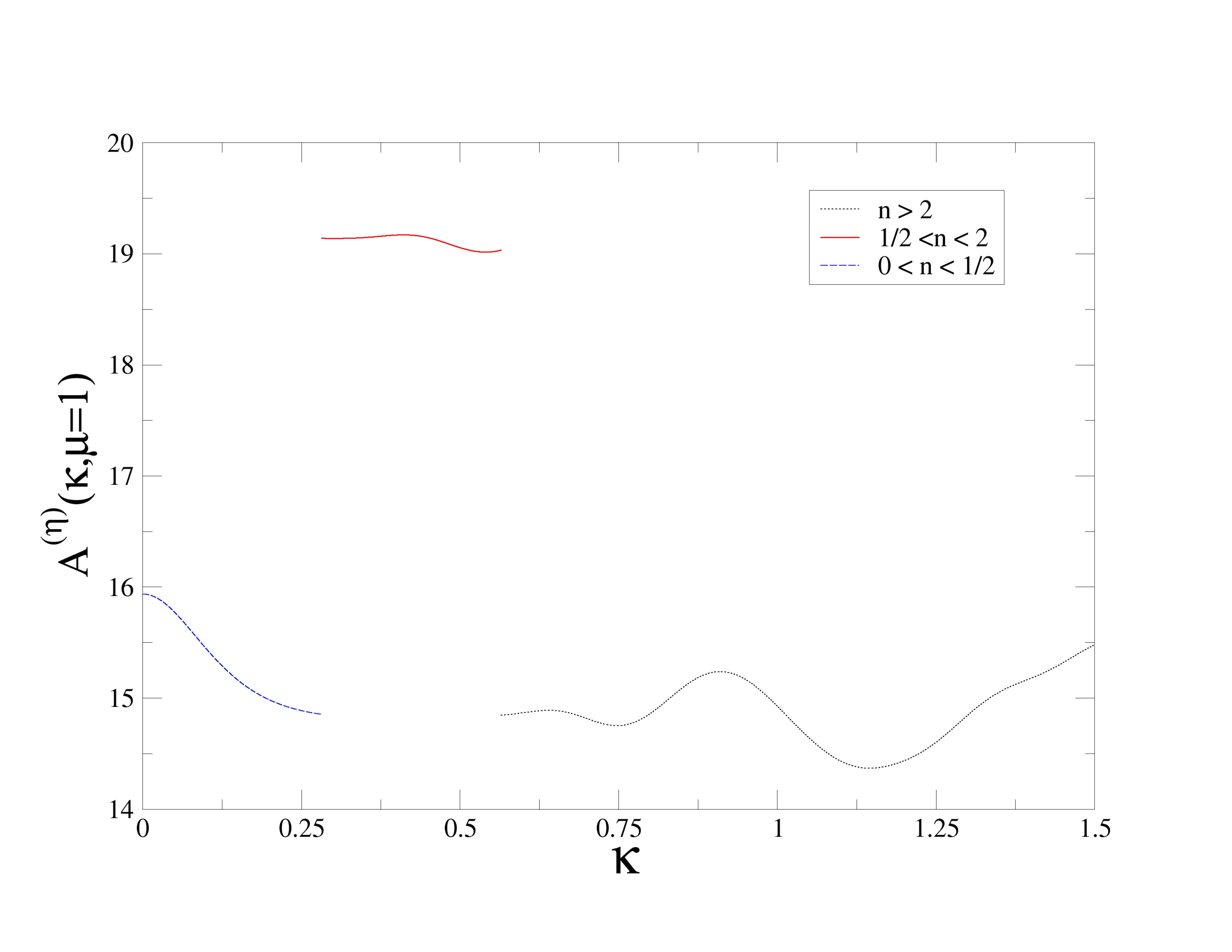}
\caption{Variation of the DRG flow integral $A^{(\eta)}_{\mu}(n;\kappa)$ with renormalized force $\kappa$ for $\mu=1$. The discontinuity observed between $0.25<\kappa<0.5$ is a signature of the phase transition to the KPZ phase.
\label{fig_mu1cosAgamma}}
\end{figure}
\end{center}

\begin{center}
\begin{figure}[tbp]
\includegraphics[height=10.0cm,width=12.0cm]{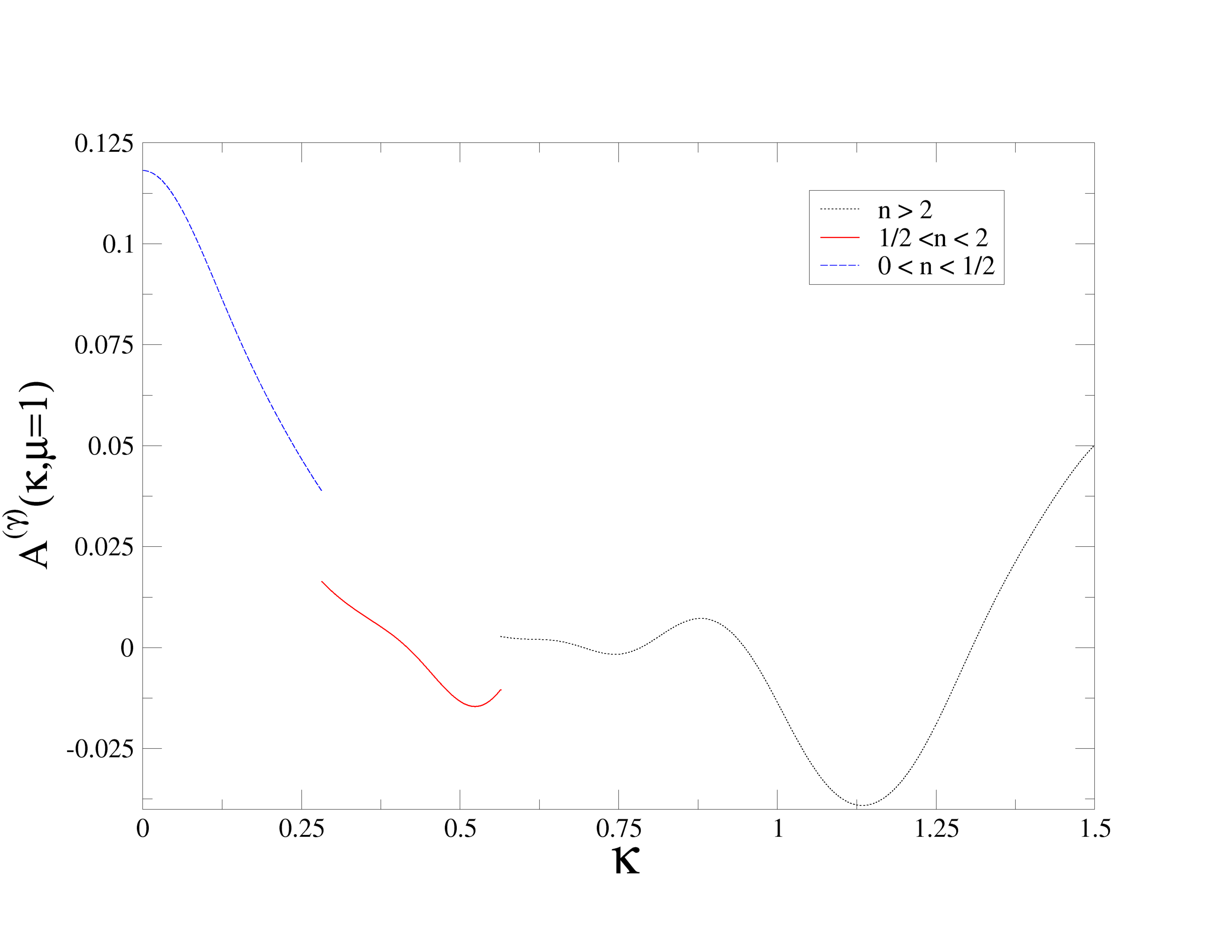}
\caption{Variation of the DRG flow integral $A^{(\gamma)}_{\mu}(n;\kappa)$ with renormalized force $\kappa$ for $\mu=1$. The discontinuity observed between $0.25<\kappa<0.5$ is a signature of the phase transition to the KPZ phase.
\label{fig_mu1cosAeta}}
\end{figure}
\end{center}

\begin{center}
\begin{figure}[tbp]
\includegraphics[height=10.0cm,width=12.0cm]{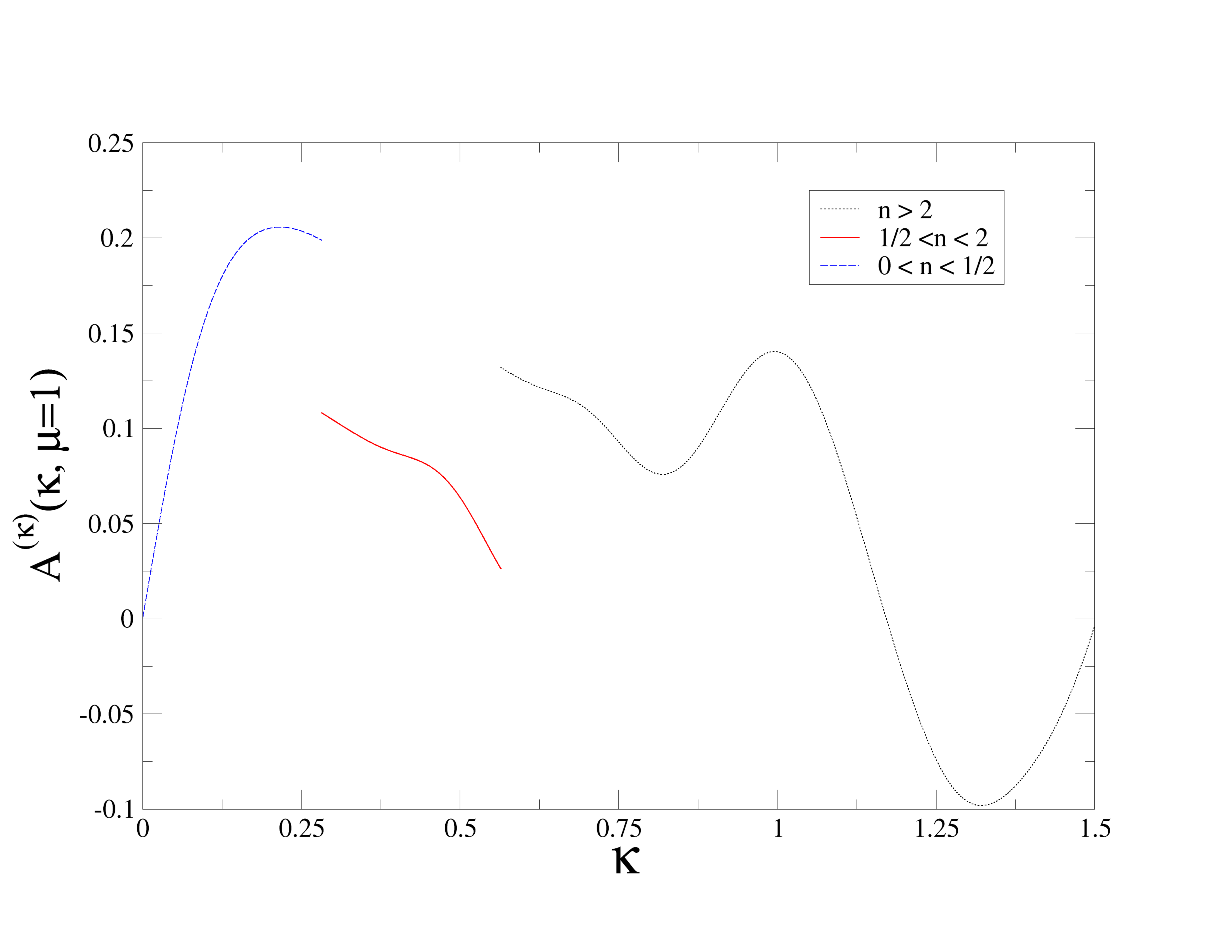}
\caption{Variation of the DRG flow integral $A^{(\kappa)}_{\mu}(n;\kappa)$ with renormalized force $\kappa$ for $\mu=1$. The discontinuity observed between $0.25<\kappa<0.5$ is a signature of the phase transition to the KPZ phase.
\label{fig_mu1cosAkappa}}
\end{figure}
\end{center}

\begin{center}
\begin{figure}[tbp]
\includegraphics[height=10.0cm,width=12.0cm]{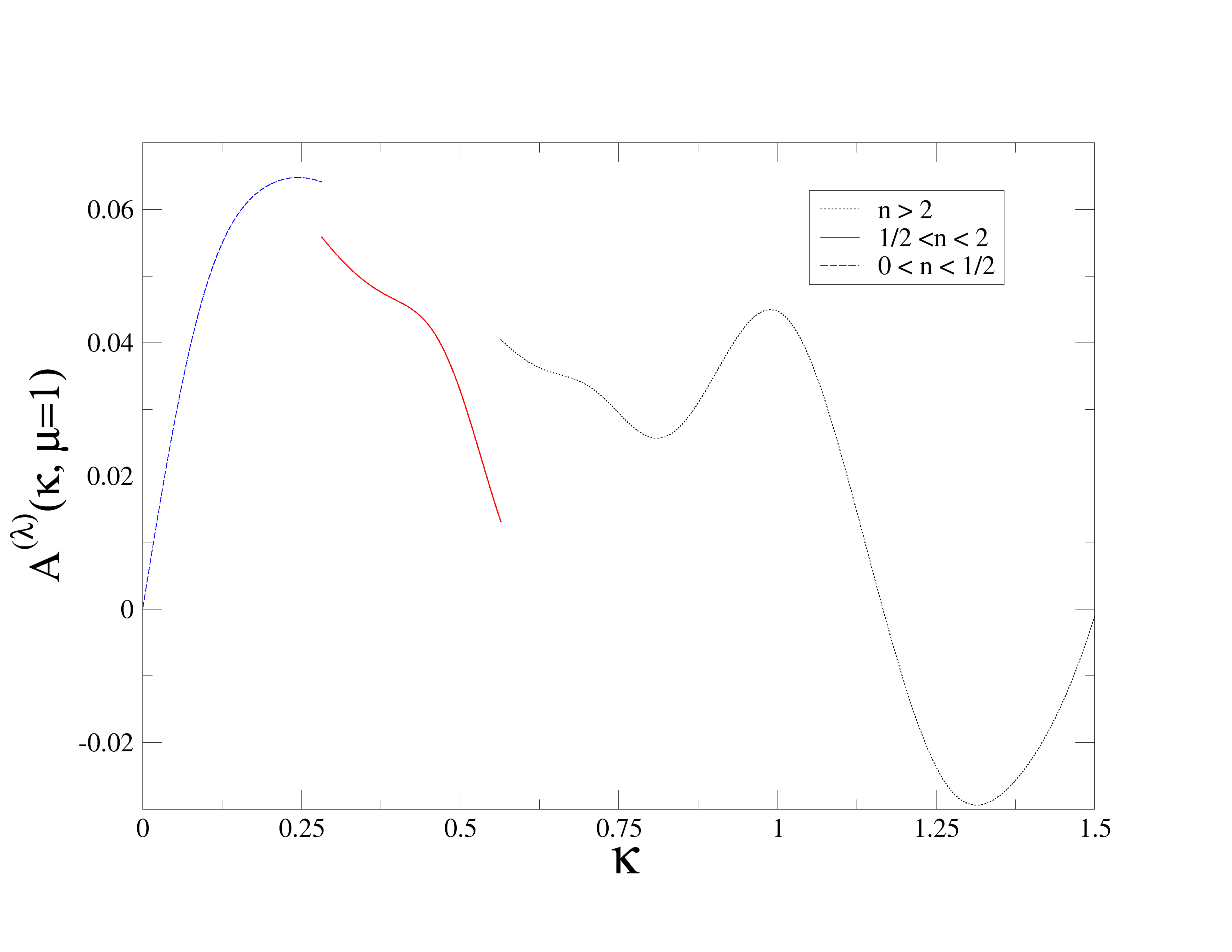}
\caption{Variation of the DRG flow integral $A^{(\lambda)}_{\mu}(n;\kappa)$ with renormalized force $\kappa$ for $\mu=1$. The discontinuity observed between $0.25<\kappa<0.5$ is a signature of the phase transition to the KPZ phase.
\label{fig_mu1cosAlambda}}
\end{figure}
\end{center}

In order to calculate the DRG flows for $\gamma$, $\eta$ and $D$, we need to 
evaluate the second ordered corrections terms $<\delta h^{(0)}\delta h^{(1)}>$ and 
{{$<\partial \delta h^{(0)}\partial \delta h^{(1)}>$}}. 
Starting from the free energy representation $\Psi$ is Eq. (9), these two-point correlation functions, accurate up to second-order in perturbation, are presented below:

\begin{subequations}
\begin{eqnarray}
\bigg \langle \delta h^{(0)}(z,t)\delta h^{(1)}(z,t) \bigg \rangle &=&-\frac {4\pi^2V_1}{a^2}\int_{-\infty}^tdt^{\prime}
\int dz^{\prime} \cos\bigg[\frac {2\pi}{a}(\bar{h}+\frac {Ft^{\prime}}{\eta})\bigg]
G_0(z-z^{\prime},t-t^{\prime}) \nonumber \\
&&\times \big \langle \delta h^{(0)}(z,t)\delta h^{(0)}(z^{\prime},t^{\prime}) \big \rangle \nonumber \\
&&+\frac {8\pi^2V_2}{a^2}\int_{-\infty}^tdt^{\prime}
\int dz^{\prime} \cos\bigg[\frac {4\pi}{a}(\bar{h}+\frac {Ft^{\prime}}{\eta})\bigg]
G_0(z-z^{\prime},t-t^{\prime}) \nonumber \\
&&\times \big \langle \delta h^{(0)}(z,t)\delta h^{(0)}(z^{\prime},t^{\prime}) \big \rangle,
\label{ncorre1}
\end{eqnarray}
and 
\begin{eqnarray}
&& {{\bigg \langle \bigg(\frac{\partial \delta h^{(0)}(z,t)}{\partial z}\bigg) \bigg( \frac{\partial \delta h^{(1)}(z',t')}{\partial z'}\bigg) \bigg \rangle}}=
{{\frac {\eta}{2\gamma}\int_{-\infty}^tdt^{\prime}
\displaystyle \int dz^{\prime} \:G_0(z-z^{\prime},t-t^{\prime})
-\left(\frac {z^{\prime}-z}{t-t^{\prime}}\right)}} \nonumber \\
&\times& \bigg\{-\frac {4\pi^2V_1}{a^2}\cos\bigg[\frac {2\pi}{a}\bigg(\bar{h}(z^{\prime},t^{\prime})+\frac {Ft^{\prime}}
{\eta}\bigg)\bigg] 
+\frac {8\pi^2V_2}{a^2}\cos\bigg[\frac {4\pi}{a}\bigg(\bar{h}(z^{\prime},t^{\prime})+\frac {Ft^{\prime}}{\eta}\bigg)\bigg]\bigg\}
\times \bigg \langle \bigg(\frac{\partial \delta h^{(0)}(z,t)}{\partial z}\bigg) \:\bigg(\frac{\partial \delta h^{(0)}(
z^{\prime},t^{\prime})}{\partial z'}\bigg)\bigg \rangle_{\delta N} \nonumber \\
&+& \frac{\eta \lambda}{\gamma}\int_{-\infty}^tdt' \int dz' \bigg(-\frac{z'-z}{t-t'}\bigg) \frac{\partial h(z',t')}{\partial z} G_0(z-z',t-t') \bigg \langle \bigg(\frac{\partial \delta h^{(0)}(z,t)}{\partial z}\bigg) \bigg(\frac{\partial \delta h^{(0)}(z',t')}{\partial z'}\bigg)\bigg \rangle _{\delta N}.
\label{ncorre2}
\end{eqnarray}
\end{subequations}

The representation in Eqs. (\ref{ncorre1}, \ref{ncorre2}) provide the detailed form of the (second-order) perturbed sine-Gordon Hamiltonian:

\begin{eqnarray}
\Psi_{SG}&\approx &-\frac {8\pi^3V_1^2T}{\gamma^2a^5}\left(\frac{a^2}{4\pi^2K^2}\right)
\Delta l\int_{-\infty}^t\frac {dt^{\prime}}
{t-t^{\prime}}\int dz^{\prime}e^{i\Lambda|z-z^{\prime}|} \times e^{-[\frac {\gamma}{2\eta}\frac {(z-z^{\prime})^2}{(t-t^{\prime})}-\frac{\gamma}
{\eta}\Lambda^2(t-t^{\prime})
-\frac {2\pi^2}{a^2}{<{[\bar{h}(z,t)-\bar{h}
(z^{\prime},t^{\prime})]}^2>}_{\delta N}]} \nonumber \\
&\times &\bigg[\frac {2\pi}{a}\bigg(\frac {\partial }{\partial t}\bar{h}(z,t)(t-t^{\prime})-
\frac 12 \frac{\partial^2 \bar{h}(z,t)}{\partial z^2}{(z-z^{\prime})}^2\bigg) \nonumber \\
&\times &\cos\bigg(\frac {2\pi}{a}\frac {F}{\eta}(t-t^{\prime})\bigg)+\bigg(1-\frac {2\pi^2}{a^2}{\bigg(
\frac{\partial \bar{h}(z,t)}{\partial z}\bigg)}^2{(z-z^{\prime})}^2\bigg) \times \sin\bigg(\frac {2\pi}{a}\frac {F}{\eta}(t-t^{\prime})\bigg)\bigg] \nonumber \\
&&-\frac {64\pi^3V_2^2T}{\gamma^2a^5}\bigg(\frac{a^2}{4\pi^2K^2}\bigg)\xi
\Delta l\int_{-\infty}^t\frac {dt^{\prime}}
{t-t^{\prime}}\int dz^{\prime}e^{i\Lambda|z-z^{\prime}|} \times e^{-[\frac {\gamma}{2\eta}\frac {(z-z^{\prime})^2}{(t-t^{\prime})}-\frac{\gamma}
{\eta}\Lambda^2(t-t^{\prime})
-\frac {8\pi^2}{a^2}{<{[\bar{h}(z,t)-\bar{h}
(z^{\prime},t^{\prime})]}^2>}_{\delta N}]} \nonumber \\
&\times &\bigg[\frac {4\pi}{a}\bigg(\frac {\partial }{\partial t}\bar{h}(z,t)(t-t^{\prime})-
\frac 12 \frac{\partial^2 \bar{h}(z,t)}{\partial z^2} {(z-z^{\prime})}^2\bigg)
 \nonumber \\
&\times &\cos\bigg(\frac {4\pi}{a}\frac {F}{\eta}(t-t^{\prime})\bigg)+\bigg(1-\frac {8\pi^2}{a^2}{\bigg(
\frac{\partial \bar{h}(z,t)}{\partial z}\bigg)}^2 {(z-z^{\prime})}^2\bigg) \times \sin\bigg(\frac {4\pi}{a}\frac {F}{\eta}(t-t^{\prime})\bigg)\bigg] .
\label{npsi5}
\end{eqnarray}

The terms in Eq. (\ref{npsi5}) that are respectively proportional to 
$\frac {\partial \bar{h}}{\partial t}$, $\partial_i\partial_j\bar{h}$ and 
$(\partial_i\bar{h})^2$ are the renormalized contributions for $\eta$, $\gamma$ 
and a new KPZ nonlinearity term $\lambda$ that automatically gets created from a
sine-Gordon potential. This drives the dynamics to the KPZ fixed point with the constant term above renormalizing the ramping force $F$. 

Combining information from Eqs. (\ref{ncorre1}, \ref{ncorre2}, \ref{npsi5}), the DRG flow equations can be outlined as follows:
%\begin{subequations}
\begin{equation}
\frac {dU_1}{dl}=(2-n)U_1,
\label{flow1}
\end{equation}
\begin{equation}
\frac {dU_2}{dl}=(2-4n)U_2,
\label{flow2}
\end{equation}
\begin{equation}
\frac {d\gamma}{dl}=\frac {8\pi^4}{\gamma a^4}n A_1^{(\gamma)}(n;\kappa)U_1^2
+\frac {64\pi^4}{\gamma a^4}n A_2^{(\gamma)}(n;\kappa)U_2^2,
\label{flow4}
\end{equation}
\begin{equation}
\frac {d\eta}{dl}=\frac{32\pi^4}{\gamma a^4}\frac{\eta}{\gamma} nA_1^{(\eta)}
(n;\kappa)U_1^2+\frac{256\pi^4}{\gamma a^4}\frac{\eta}{\gamma} nA_2^{(\eta)}
(n;\kappa)U_2^2
\label{flow5}
\end{equation}
\begin{equation}
\frac {d\lambda}{dl}=\frac {32\pi^5}{\gamma a^5}nA_1^{(\lambda)}(n;\kappa)U_1^2
+\frac {256\pi^5}{\gamma a^5}nA_2^{(\lambda)}(n;\kappa)U_2^2,
\label{flow6}
\end{equation}
\begin{equation}
\frac {dD}{dl}=\frac{32\pi^4}{\gamma a^4}\frac{D}{\gamma} nA_1^{(\eta)}
(n;\kappa)U_1^2+\frac{256\pi^4}{\gamma a^4}\frac{D}{\gamma} nA_2^{(\eta)}
(n;\kappa)U_2^2 + \frac{1}{4\pi}\frac{D^2 \lambda^2}{\gamma^3}
\label{flow7}
\end{equation}
\begin{equation}
\frac {d K}{dl}=2K-\frac {8\pi^3}{\gamma a^3}nA_1^{(K)}(n;\kappa)U_1^2
-\frac {64\pi^3}{\gamma a^3}nA_2^{(K)}(n;\kappa)U_2^2 + \frac{D}{2\pi \eta \gamma}\lambda,
\label{flow8}
\end{equation}
%\end{subequations}

where $U_\mu=V_\mu/\Lambda^2$ ($\mu$ = 1,2), $K=F/\Lambda^2$, $\bar{\rho}=\Lambda \rho$, $x= \frac {\gamma(t-t^{\prime})}{\eta \rho^2}$, $D=\frac {a^2}{4\pi^2K^2}\eta T$,
$n=\frac{2\pi D}{\gamma a^2}(\frac {a^2}{4\pi^2K^2})$ and $\kappa=\frac {2\pi K}{a\gamma}$.

The functional forms of the DRG flow integrals $A^{(\gamma)}(n;\kappa)$, $A^{(\eta)}(n;\kappa)$, $A^{(\lambda)}(n;\kappa)$ and $A^{(K)}(n;\kappa)$ are defined as follows:
%\begin{subequations}
\begin{eqnarray}
A^{(\gamma)}_{\mu}(n;\kappa) &=& \displaystyle \int_0^{\infty}\frac {dx}{x}\int_0^{\infty}d\bar{\rho}\:
\bar{\rho}^2e^{i\bar{\rho}}\cos\left(\frac {2\pi \mu}{a}\frac {Kx\bar{\rho}^2}{\gamma}\right)\:e^{-f_{\mu}}, \\
A^{(\eta)}_{\mu}(n;\kappa) &=& \frac{x}{\bar{\rho}^2}A^{(\gamma)}_\mu(n;\kappa), \\
A^{(\lambda)}_\mu (n;\kappa) &=& \displaystyle \int_0^{\infty}\frac {dx}{x}\int_0^{\infty}d\bar{\rho}\:
\bar{\rho}^2e^{i\bar{\rho}}\sin\left(\frac {2\pi \mu}{a}\frac {Kx\bar{\rho}^2}{\gamma}\right)\:e^{-f{\mu}}, \\
A^{(K)}_\mu (n;\kappa) &=& \frac{1}{\bar{\rho}^2}A^{(\lambda)}_\mu (n;\kappa), 
\end{eqnarray}
%\end{subequations}

where the components can be estimated from the following identities
\begin{eqnarray}
f_1 &=& \frac {1}{2x}+x\bar{\rho}^2+\frac {2\pi D}{\gamma a^2}\chi(\bar{\rho},x), \nonumber\\
f_2 &=& \frac {1}{2x}+x\bar{\rho}^2+\frac {8\pi D}{\gamma a^2}\chi(\bar{\rho},x), \nonumber\\
\chi(|z-z'|,|t-t'|) &=& \frac {\pi\gamma}{\eta T} <[\bar{h}(z,t)-\bar{h}(z^{\prime},t^{\prime})]^2>_{\bar{N}}, \nonumber\\
<[\bar{h}(z,t)-\bar{h}(z^{\prime},t^{\prime})]^2>_{\bar{N}} &=& \frac{\eta T}{\pi \gamma}\displaystyle \int_0^1dk\big[\sqrt{2(1-\cos{k|z-z'|})}\big]\:e^{-\frac{\gamma}{\eta}k^2(t-t')},
\end{eqnarray}

where $\mu=1,2$. The KPZ fixed point relates to the generation of the renormalized term $(\vec{\nabla}h)^2$
as a result of the dynamical evolution of the model. Here $n=\frac {D}{2\pi\gamma K^2}$; for $n>2$, both $U_1$ and $U_2$ decay
to zero while for $\frac 12<n<2$, only $U_2$ decays to zero while $U_1$ keeps
growing. The case for $n=2$ is most interesting in that it leads to a strong coupling fixed point where the $U_1$ flow diverges, an attribute that gets subsequently reflected in the phase evolution of all relevant quantities like $\gamma$, $\eta$, $D$, $K$ and the renormalized parameter $\lambda$. This indicates a phase transition from the $N^{*}$ phase to the 
SmC$^{*}$ phase. Unlike in the N$^{*}$ phase, for each set of $\gamma$, 
$\eta$, $\lambda$, $D$ and $K$ flows, there is a push-pull mechanism in action.
For $\frac 12<n<2$, we predict that the system will be in the N$^*$
phase while in the other regions they will flow in the SmC$^*$ phase.
This observation supports the expermental observations \cite{patel86,biradar96,biradar00} of the 
N$^{*}$-SmC$^{*}$ phase transition. Figures 1-4 highlight these phase transitions from the $N^{*}$ phase to the 
SmC$^{*}$ phase, focusing on the convergence properties of the flow integrals. The double discontinuities at the points $\kappa=0.25$ and $\kappa\sim 0.56$ indicate these phase transition points.

\section*{Numerical solution}
%\label{numerics}

In order to analyze the detailed temperature dependence of the model, an allusion to the fact that the noise strength $D_0$ could be related to the Brownian mobility and hence is proportional to $k_B T$, where $k_B$ is the Boltzmann's constant and $T$ is the {{\enquote{effective non-equilibrium temperature} \cite{Jarzynski2011, Bustamente2005, Seifert2012, Derrida2007}}} of the system (discussed earlier), we resorted to  numerical simulation. 

{{For the actual numerical integration of our 1+1 dimensional model, we used the \enquote{forward Euler} discretization scheme with $\Delta t={10}^{-4}$ and $\Delta x=1$. The Laplacian term in Eq. (\ref{lang5}) was discretized within the iterative nearest-neighbor representation, followed on by next-nearest-neighbor, etc. through a recursion loop. In order to ensure dynamic stationarity, the system was dynamically evolved through ${10}^5$ time steps. Diffusion constant $\gamma=1$ and lattice spacing $a=1$ were the only fixed parameter values chosen (without any loss of generality). In order to ensure the parameter independence of the fixed points, thereby confirming existence of proper universality classes, we ran simulations over all possible combinations of the following parameter values: $\dfrac{V_1}{V_2}=0.1,\:1,\:10$,  $\eta=1,10$, $F=1,\:10,:100$, low noise strength $D=0.1,\:0.5,\:1$. Plots are shown only for a few specific choice of these parameter values but the choices as such are arbitrary, since both qualitative and quantitative predictions of the exponent values could be confirmed to remain independent of the choice of parameters. As expected, for even larger values of the noise strength, as also for overly large choices of $\eta$, the results trivially converged to the Brownian model or to the \enquote{flat, non-rough} stationary limit respectively. Also, we avoided combinations of parameter values (tested through simple dimensional scaling of terms) for which the diffusion, sine-Gordon and noise terms were not sufficiently competitive, as the prediction from the model relies on this competition. For individual asymptotes, e.g. $\gamma>>V_1,\:V_2$, or the reverse, the results converged to known fixed points.

In order to clarify the finite sized dependence of this model, lattice sizes of 1000, 10,000 and 100,000 points were used. From our results, we can now safely claim that the universality classes themselves are unaffected by finite sized corrections. The simulation results provide a deeper insight into the dynamical phase transition, the full range of which has hitherto remained elusive to even most updated experimental attempts \cite{patel86,biradar96,biradar00}. 

The scaling regimes depicted in Figures 5 and 6 represent the fixed points defining the corresponding universality classes. The solid line shows a convergence to the value $\alpha\sim0.5,\:\beta\sim1/3$ while the dotted line scales as $\alpha\sim1/3,\:\beta\sim1/3$. It may be noted that the results shown are block ensemble averages over ${10}^4$ time realizations in each case, and as such are most accurate. Each data point shown is actually a representation of this time/ergodic sampling. 
}}

\subsection*{KPZ phase confirmation}

Essentially, we are calculating two two-point correlation functions, the spatial correlation function $C_x=<{[h(x+z,t)-h(z,t)]}^2>$ and  the temporal correlation function $C_t=<{[h(z,t+\tau)-h(z,\tau)]}^2>$. The universality class as such can be confirmed from their respective scaling exponents, the \enquote{roughness exponent} $\alpha$, the \enquote{growth exponent} $\beta$ and the \enquote{dynamic exponent} $\gamma$. As is well known \cite{barabasi,krug}, due to self-affine scaling, such systems can be uniquely expressed by any two of these exponents. e.g. $\alpha$ and $z$, where $\beta=\dfrac{\alpha}{z}$. 

For large forcing, that is for reasonably large values of $F>1$, the trajectories converge to a KPZ fixed point characterized by $\alpha\sim0.5$ and $z\sim1.5$ \cite{kpz,barabasi}. We may add that $F=1$ is not a special parameter value though. This result confirms the experimental finding of \cite{sano1,sano}; {{the term $F_0$ in our model corresponds to the steady external voltage that was applied to the thin nematic liquid crystal layer in these references.

A conventional strategy employed in some previous works \cite{nozieres,rost} while not availed in others \cite{akc} is the incorporation of the KPZ term in the dynamic renormalization structure right from the onset. For the purpose of this manuscript, it is believed to be an incorrect strategy in view of the fact that the later numerical simulation shows the evidence of non-KPZ phases as well. If a KPZ term is added in the starting model itself, this would trivialise the importance of the KPZ convergence for large values of the external forcing. 

\begin{center}
\begin{figure}[tbp]
\includegraphics[height=10.0cm,width=12.0cm]{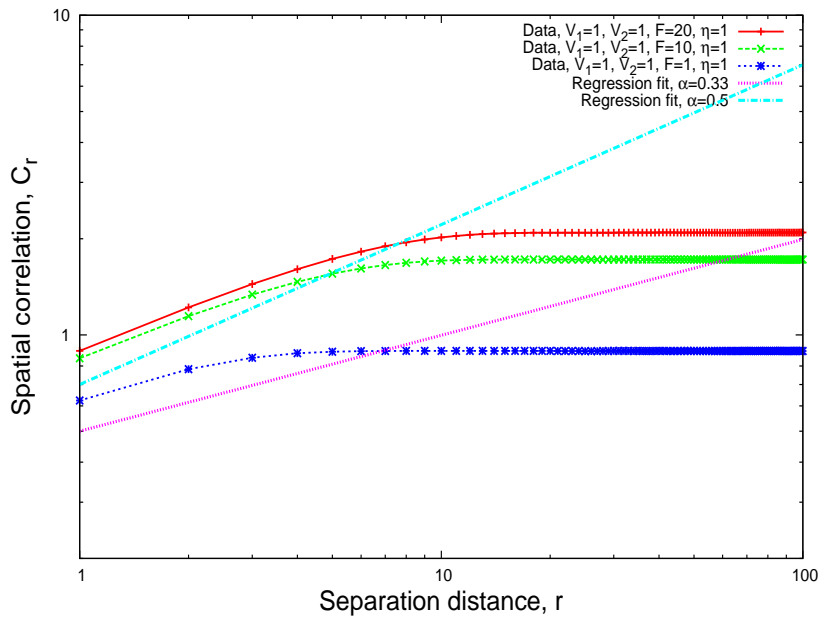}
\caption{
Spatial correlation function $C_x$ plotted against the spatial separation $x$ in the loglog scale for $V_1=1, \:V_2=1, \:\eta=1$, respectively for $F=1,\:10,\:20$, as shown in the plots. The results show a remarkable phase transition from the \enquote{subdiffusive} phase defined by $\alpha \sim 0.33,\:\beta \sim 0.33$ to a KPZ \cite{kpz} phase defined by $\alpha\sim 0.5,\:\beta\sim 0.33$. As the ramping force $F$ increases from a relatively low ($F=1$) to high values ($F=10,\:20$), the system crosses over from a subdiffusive universality class to a KPZ universality class. 
\label{fig_Fig5}}
\end{figure}
\end{center}

\begin{center}
\begin{figure}[tbp]
\includegraphics[height=10.0cm,width=12.0cm]{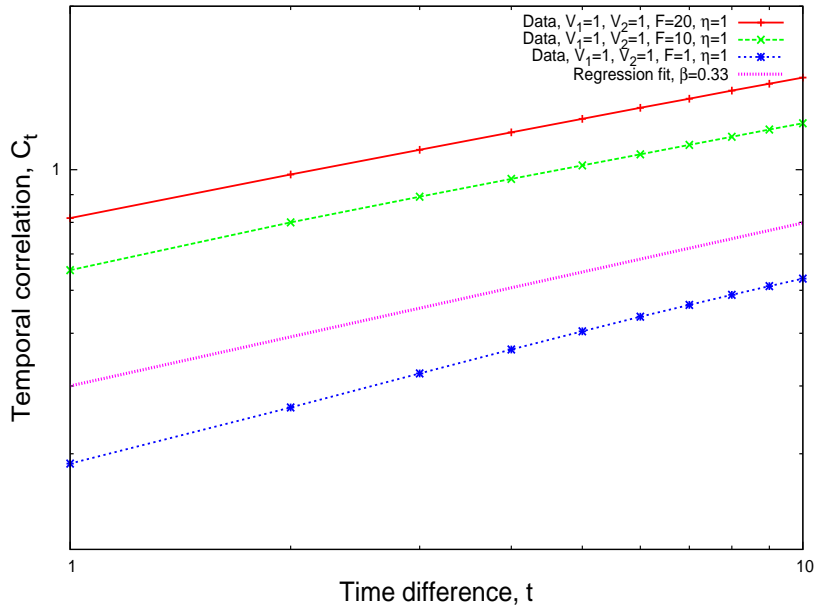}
\caption{
Temporal correlation function $C_t$ plotted against time difference $t$ in the loglog scale for $V_1=1, \:V_2=1, \:\eta=1$, respectively for $F=1,\:10,\:20$, as shown in the plots. All gradients converge to the fixed value $\beta \sim 0.33$. The growth exponent is clearly unaffected by the ramping force $F$.  
\label{fig_Fig6}}
\end{figure}
\end{center}

}}

%\begin{center}
%\begin{figure}[tbp]
%\includegraphics[height=10.0cm,width=12.0cm]{Fig5.jpg}
%\caption{Spatial correlation function $C_x$ plotted against the spatial separation $x$ in the loglog scale for $V_1=1, \:V_2=1, \:\eta=1$, respectively for $F=1,\:10,\:20$, as shown in the plots. The results show a remarkable phase transition from the \enquote{subdiffusive} phase defined by $\alpha \sim 0.33,\:\beta \sim 0.33$ to a KPZ \cite{kpz} phase defined by $\alpha\sim 0.5,\:\beta\sim 0.33$. As the ramping force $F$ increases from a relatively low to high value, the phase changes from a subdiffusive universality class to a KPZ universality class.
%\label{fig_spacecorr}}
%\end{figure}
%\end{center}

%
%\begin{center}
%\begin{figure}[tbp]
%\includegraphics[height=10.0cm,width=12.0cm]{Fig6.jpg}
%\caption{Temporal correlation function $C_t$ plotted against time difference $t$ in the loglog scale for $V_1=1, \:V_2=1, \:\eta=1$, respectively for $F=1,\:10,\:20$, as shown in the plots. All gradients converge to the fixed value $\beta \sim 0.33$. The growth exponent is clearly unaffected by the ramping force $F$.
%\label{fig_timecorr}}
%\end{figure}
%\end{center}

%,vallerien90,biradar92,bawa93,biradar96,biradar96a,biradar99,asao99,biradar00}. 

\subsection*{The \enquote{Hidden} phase: Subdiffusive to KPZ}
In a remarkable finding, as $F$ decreases from a high value $F\ge 10$ to a relatively lower value e.g. $F=1$ and below, the system dynamically converges to a \enquote{subdiffusive} (SD) universality class characterized by ($\alpha\sim0.33,\:\beta\sim0.33,\:z\sim1$). As $F$ is again ramped higher, the trajectories cross over to a KPZ fixed point \cite{kpz} characterized by ($\alpha\sim0.5,\:\beta\sim0.33,\:z\sim1.5$). Figures 5 and 6 highlight these scaling features and confirms the crossover from the SD to the KPZ phase, and vice versa.

The emergence of an SD phase can be understood from the previously estimated RG flows which define two {{\enquote{jump} transition points}} at $n=\dfrac{D \Lambda^2}{2\pi \gamma F^2}=\dfrac 12$ and $n=\dfrac{D \Lambda^2}{2\pi \gamma F^2}=2$ respectively, as corroborated by the discontinuities at the corresponding points in all the flow integral plots shown in Figs. (1-4). As $F$ increases, $n\to\frac 12$ which identifies the KPZ phase in the phase integral plots while with decreasing $F$, the $n\to2$ fixed point is approached. Fig. 5 above clearly shows how the roughness exponent $\alpha$ (defined in \cite{kpz}) changes from a smoother SD phase, identified by $\alpha\sim 0.33$ to a coarser KPZ phase, characterized by $\alpha\sim 0.5$, thereby leading to a dynamic exponent value of $z \sim 1.5$ in the KPZ phase. {{These results could be compared with the preroughening-roughening transitions previously noted in \cite{nijs1989,park1993}.}} Fig. 5 confirms the values of the \enquote{roughness exponent} \cite{kpz} over a range of $F$ values. The KPZ-growth exponent value remains unchanged for both SD and KPZ dynamics, as is shown in Fig. 6.

The crossover from the SD phase to the KPZ phase has concurrently appeared over this entire parametric range. A crucial feature of this report concerns the prediction of this \enquote{hidden phase} and its subsequent transition to the Kardar-Parisi-Zhang (KPZ) phase, as previously reported in experiments \cite{sano1,sano}. We believe that in this ground breaking experimental study, the SD phase remained elusive due to an inappropriate choice of parameters, attributed to a lack of mechanism to choose parameters over such a wide parameter space. Our model study now presents a structure to arrive at this parameter set whereby such a \enquote{hidden} SD phase could be experimentally verified, that at the present, seems to be characterized only by a KPZ dynamics. We have checked the veracity of this prediction for other values of ($V_1,\:V_2$), within the range $1<V_i<10$ ($i$=1, 2). 

{{ From the perspective of quantum spin chains \cite{nijs1989} as also in roughening transition, similar multi-phase properties have previously been reported \cite{park1993}, in which the transition from the $n=1/2$ to the $n=2$ phase respectively represented a half-integral spin to an integral spin conformation transition, or a pre-roughening to a roughening transition (or a cross-over). These studies also highlight the importance of possible long-ranged phase correlations, and hence spatially controlled transitions \cite{akc1999}, that could be studied in the present context as well.}}

\section*{Conclusions}
Using a complementary combination of analytical (dynamic renormalization group) and numerical techniques, we explore the nonequilibrium scaling features of a stochastically perturbed smectic-C${}^{*}$ liquid crystal that on the one hand confirms known experimental results (prediction of a KPZ phase), while paradigmatically, unearths a \enquote{hidden} phase (subdiffusive phase) that has eluded experimentalists thus far.
 
The prediction of the KPZ universality class from our model, in 1+1 dimensions, validates a recent experimental work by 
Takeuchi and Sano \cite{sano} where they studied the scale-invariant 
fluctuations of growing interfaces in nematic liquid crystal turbulence.
As a perturbed phase, the N$^*$ and SmC$^*$ phases too are expected to behave 
identically. From a general perspective, for experiments involving more than one spatial dimension, the remit of this model could be easily extended as in \cite{akc}. Even for such high-dimensional phases, our model predicts the emergence of a KPZ universality class. Figures 1-4 affirm the transition to the KPZ fixed points noted in these experiments.

A remarkable feature of our numerical analysis is the identification of a hitherto unexplored subdiffusive \enquote{smoother} phase characterized by critical exponent values ($\alpha=\frac 13,\:\beta=\frac 13,z=\frac{\alpha}{\beta}=1$). We find that this phase flows over to a KPZ phase with vamped up external forcing ($F$) but otherwise remains dynamically preserved. In other words, this is not a transient feature of this model analysis and should be observable from experiments. We have provided a range of parameter values whereby such a phase could be traced. 

Here, we have studied the impact of stochastic fluctuations in modifying the spatiotemporal properties of ferroelectric SmC$^*$ liquid 
crystals in the presence of an electric field. Our model predicts 
a first order N$^*$-SmC$^*$ phase transition, and thereby offers the first theoretical 
ramification of the experimental observations of a KPZ phase as in \cite{sano1,sano}, with others \cite{patel86,biradar96,biradar00} following suit. Our approach extends the remit of the experimental finding of the emergence of a KPZ universality class, showing that other universality classes could also appear, while identifying parameter regimes to look for the same. The implication of finite sized impurities and their contribution in related (likely non-universal) dynamical regimes are likely to enthuse a new generation of material science research, routed in liquid crystals.

\end{document}